\documentclass{iaus}
\usepackage{graphicx}

\title[The rotational evolution of young low mass stars ] 
{The rotational evolution of young low mass stars}

\author[J. Bouvier]   
{J\'er\^ome Bouvier}

\affiliation{Laboratoire d'Astrophysique, Observatoire de Grenoble,
Universit\'e Joseph Fourier, \break B.P.53, 38041 Grenoble, Cedex 9, France
\break email: jbouvier@obs.ujf-grenoble.fr}

\pubyear{2007}
\volume{243}  
\pagerange{??--??}
\date{?? and in revised form ??}
\jname{Star-Disk Interaction in Young Stars}
\editors{J. Bouvier \& I. Appenzeller, eds.}
\begin{document}

\maketitle

\begin{abstract}
Star-disk interaction is thought to drive the angular momentum evolution of
young stars. In this review, I present the latest results obtained on the
rotational properties of low mass and very low mass pre-main sequence
stars. I discuss the evidence for extremely efficient angular momentum
removal over the first few Myr of pre-main sequence evolution and describe
recent results that support an accretion-driven braking mechanism. Angular
momentum evolution models are presented and their implication for accretion
disk lifetimes discussed.

\end{abstract}

\section{Introduction}

Star-disk interaction in young systems involves mass and angular momentum
transfer. Mass transfer manifests itself by accretion onto the star and
wind/jet outflows while angular momentum transfer is thought to drive the
rotational evolution of young stars and is also seen in jet rotation
(cf. T. Ray's review in this volume). The purpose of this review is to
address the following issues : i) what do observations tell us about the
rotational evolution of young stars, ii) what is the impact of star-disk
interaction on angular momentum evolution during the pre-main sequence, and
iii) how successful are models in accounting for the observed rotational
evolution of young stars ?

Section 2 describes the rotational properties of low mass (0.2-1.0
M$_\odot$) pre-main sequence stars. Recent results include the
determination of rotation periods for hundreds of stars in young clusters
spanning an age range from $\leq$1~Myr to the zero-age main sequence and
beyond. For the first time, these results provide a relatively well sampled
age sequence, from which the rotational evolution of young stars can be
read.

Since the early 90's, it is commonly thought that the star-disk
interaction is somehow reponsible for the low rotation rate of young
stars. Section 3 discusses the issue of accretion-related angular
momentum loss, a process which should manifest itself by accreting
stars being, on average, slower rotators than non accreting
ones. Conflicting evidence for such a connection has been reported and
is revisited here in the light of new results.

Section 4 provides a brief status of current angular momentum evolution
models. Most models rely on the same simplified assumptions and still await
to be upgraded to an actual physical description of the angular momentum
loss processes acting during the pre-main sequence. The implications of
these models for the lifetime of accretion disks are discussed.

\section{The rotational periods of young low mass stars}

In the last 10 years, rotational periods have been measured for
hundreds of low mass pre-main sequence stars in various star forming
regions and young open clusters. Rotational periods are derived by
detecting a periodic component in the star's photometric variability,
which results from the modulation of the star's brightness by surface
spots. While this observational approach is time consuming, requiring
the photometric monitoring of hundreds of stars over a timescale of
several weeks, it is also far superior to spectroscopic $v\sin i$
measurements as the photometrically-derived rotational period is not
affected by inclination effects. Moreover, the star's angular
velocity, a key physical parameter of star-disk interaction models,
relates directly to the rotational period ($\Omega =2\pi /P$), without
having to resort to the poorly known stellar radius. These properties
have been prime motivations for lauching large scale photometric
monitoring campaigns targetting star forming regions with the hope to
derive the rotational period distribution of statistically significant
samples of pre-main sequence stars at various stages of their
evolution.

The first large scale photometric monitoring campaign by Choi \&
Herbst (1996) revealed the peculiar {\it bimodal\/} distribution of
periods for young stars in the Orion Nebulae Cluster (ONC). The
distribution of periods at an age of about 1~Myr exhibits two peaks,
one located at a period of about 8 days, the other at a period of
about 2-3 days, with approximately twice as many slow rotators as 
fast ones.  While this result was disputed a few years later (Stassun
\etal\ 1999), the bimodal distribution was eventually confirmed, with
peaks near 2 and 8 days, for ONC stars more massive than
0.25~M$_\odot$ (Herbst \etal\ 2002). Interestingly, the same study
also showed that the bimodality does not extend to lower mass stars
whose period distribution exhibits a single peak at fast rotation with
a mean period of about 2 days. This was the first hint at possibly
different rotational properties between low mass (0.3-1.0~M$_\odot$)
and very low mass ($\leq$0.3~M$_\odot$) young stars.

\begin{figure}
\centering
 \includegraphics[width=0.8\textwidth]{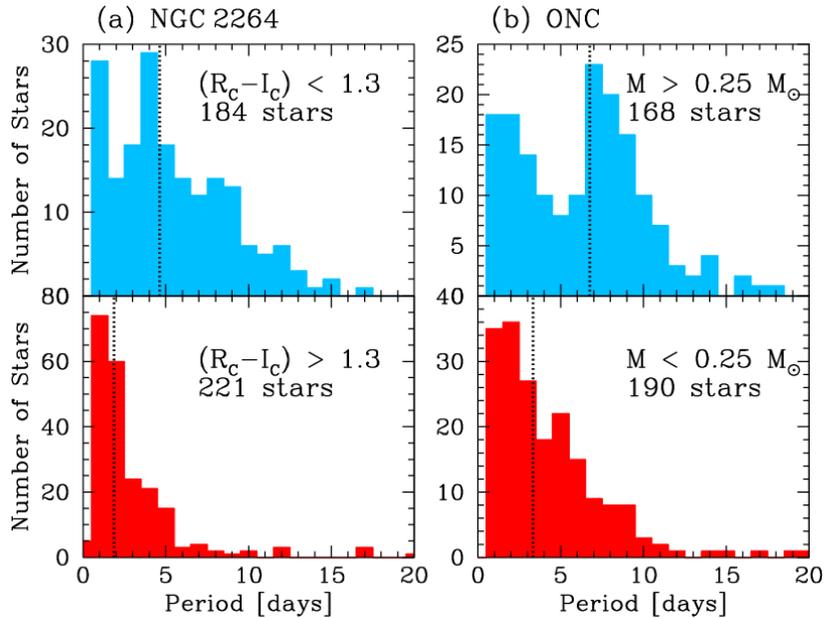}
  \caption{The rotational period distributions of low mass ({\it top})
  and very low mass ({\it bottom}) stars in the $\sim$2-4~Myr NGC 2264
  cluster ({\it left}) and in the $\sim$1~Myr Orion Nebulae Cluster
  ({\it right}). From Lamm \etal\ (2005)}\label{lamm}
\end{figure}

One of the first comparative studies between 2 clusters of different
ages came from the measurement of hundreds of rotational periods for
low mass stars in NGC~2264 (Lamm \etal\ 2005). With an estimated age
of 2-4~Myr, NGC~2264 is slightly older than ONC, thus allowing a first
attempt to trace the evolution of rotation at the start of pre-main
sequence evolution. The distribution of periods for ONC and NGC~2264
are shown in Figure~\ref{lamm} for low mass and very low mass stars, with a
dividing line around 0.25~M$_\odot$. In each mass bin, the rotational
distributions for the 2 clusters have a similar shape, bimodal for low
mass stars and single-peaked for very low mass stars. However, on a
timescale of a few Myr from ONC to NGC~2264, the peaks of the
distributions have shifted towards shorter periods, indicative of
stellar spin up by a factor of about 1.5 to 2. Within the
uncertainties on the age of the 2 clusters and assuming that the
initial period distributions were similar in both clusters, this
amount of spin up is consistent with stellar contraction and angular
momentum conservation (Lamm \etal\ 2005). While most stars appear to
spin up between ONC and NGC~2264, a tail of slow rotators remains,
which suggests efficient braking for a fraction of low mass stars
during the first 2~Myr of their evolution.

\begin{figure}
\centering
 \includegraphics[width=1.0\textwidth,angle=270]{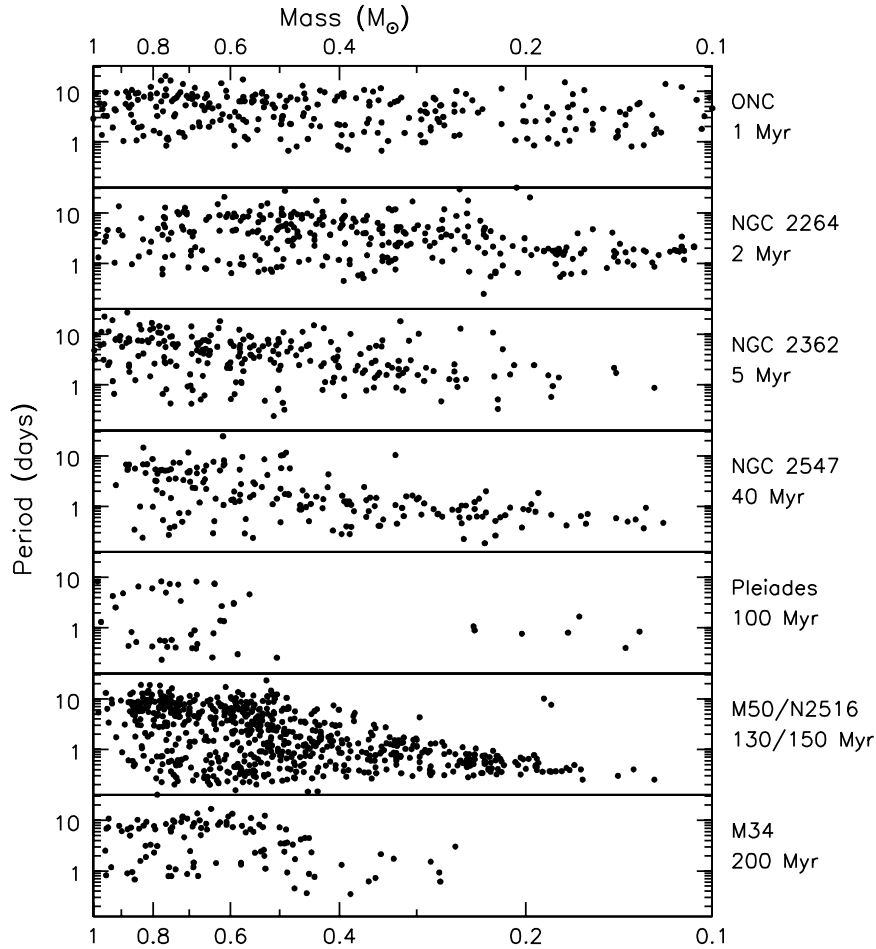}
  \caption{The rotational period distribution of low mass and very low
  mass stars in various clusters in the age range from 1 to
  200~Myr. Note the evolution of the shape of the period distribution
  as a function of time, especially at very low masses. Adapted from
  Irwin \etal\ (2007b) and references therein.}\label{monitor}
\end{figure}

A major leap in the study of angular momentum evolution of young stars
recently came from the derivation of hundreds of rotational periods for low
mass and very low mass stars in various clusters spanning an age range from
5 to 200~Myr. The Monitor project (Aigrain \etal\ 2007; Irwin et al. 2007a)
aims at detecting eclipsing binaries and planetary transits in young low
mass stars. As a by-product of intense photometric monitoring campaigns,
precise rotational periods are derived for the young populations of these
clusters (Irwin \etal\ 2006, 2007b). The period distributions derived for
stars in the mass range 0.1-1.0~M$_\odot$ are shown in Figure~\ref{monitor}
over the age range 1-200~Myr.  The shape of the period distribution evolves
drastically over the first 40~Myr. Starting from ONC at 1~Myr, the very low
mass stars (M$\leq$0.25~M$_\odot$) appear to continuously spin up, from a
median initial period of 2-3 days at 1~Myr to a median period of 0.5-0.7
days at 40~Myr and even shorter at 150~Myr. The rapid convergence of very
low mass stars towards fast rotation, with no slow rotators left, is quite
remarkable indeed.

As a group, higher mass stars (M$\geq$0.5~M$_\odot$) also tend to spin
up during PMS evolution, though not as much as very low mass stars,
with a median initial period of about 6 days at 1~Myr shortening to
about 3 days at 40~Myr. However, the initial bimodal distribution of
periods seen for stars in this mass range remains visible over the
whole PMS evolution. A fraction of initially slowly rotating stars,
with periods of 8-15 days at 1~Myr, retain similar periods up to 5~Myr
and experience only mild spin up later on, with periods in the range
5-10 days at 40~Myr. Meanwhile, the initially fast rotators, with
periods of 1-3 days at 1~Myr, spin up continuously with the shortest
periods decreasing from 0.9d at 1~Myr, to 0.5d at 5~Myr and 0.2d at
40~Myr. 

Recent results thus converge in indicating a somewhat different
rotational evolution between low mass and very low mass stars during
the pre-main sequence. Most very low mass stars
(M$\leq$0.25~M$_\odot$) start their evolution as fast rotators and
spin up continuously as they descend their Hayashi tracks and approach
the ZAMS. In contrast, a significant fraction of low mass stars
(0.3-1.0~M$_\odot$) experience milder spin up during their pre-main
sequence evolution. Indeed, some appear to evolve at constant angular
velocity for the first 5~Myr at least (see also Rebull et
al. 2004). Clearly, a very efficient brake must be at work to extract
angular momentum in these young stars in order to prevent them from
spinning up in spite of stellar contraction. 

\section{Is PMS braking related to the accretion process ? }

On the main sequence, solar-type stars are braked at a low pace by their
magnetized winds. Even though magnified versions of solar-type winds
probably exist in the magnetically active low mass PMS stars, the
associated braking timescale is much longer than the Kelvin-Helmotz
contraction timescale (Bouvier \etal\ 1997). As a result, these winds
cannot prevent the star from spinning up as it contracts towards the ZAMS
(see Matt, this volume).

As an alternative, following models developped for X-ray binaries,
K\"onigl (1991) suggested that the magnetic star-disk interaction
could regulate the angular momentum of the star. This idea has since
been developped in a variety of MHD models where angular momentum is
extracted from the star by the magnetic field and carried away by the
disk or by an accretion-driven wind (see the contributions by Shu,
Fendt, Romanova, and Ferreira in this volume). In these models, the
star is thus braked as long as it accretes from its disk. 
 
Irrespective of the specific underlying physical model (X-wind, disk
locking, accretion-driven disk winds or stellar winds), any
accretion-related angular momentum loss process should reveal itself as
accreting stars being, on average, slower rotators than non accreting
ones. Such a relationship between rotation and accretion has therefore been
actively searched for. A first hint that such a correlation may exist was
reported for a limited sample of T Tauri stars in Taurus with known
rotational period and IR excess (Edwards et al., 1993). A larger scale
study of ONC stars however revealed no such correlation (Stassun \etal\
1999).

One difficulty in searching for rotation-accretion connection, besides
statistical robustness, is to identify an unambiguous diagnostics of
accretion onto the star. The often used near-IR and mid-IR excesses
are somewhat ambiguous in this respect as they can arise from passive,
i.e., non accreting, disks. The UV continuum excess is a more direct
measurement of accretion onto the star but has been measured for too
few young stars to be useful. H$_\alpha$ line emission is a proxy of
accretion onto the stellar surface, as long as it can be shown that
the line flux or width exceeds the chromospheric emission
component. In some regions with strong nebular H$_\alpha$ emission,
like ONC, the interpretation of the H$_\alpha$ line is however not
straighforward.

\begin{figure}
\centering
 \includegraphics[width=0.8\textwidth]{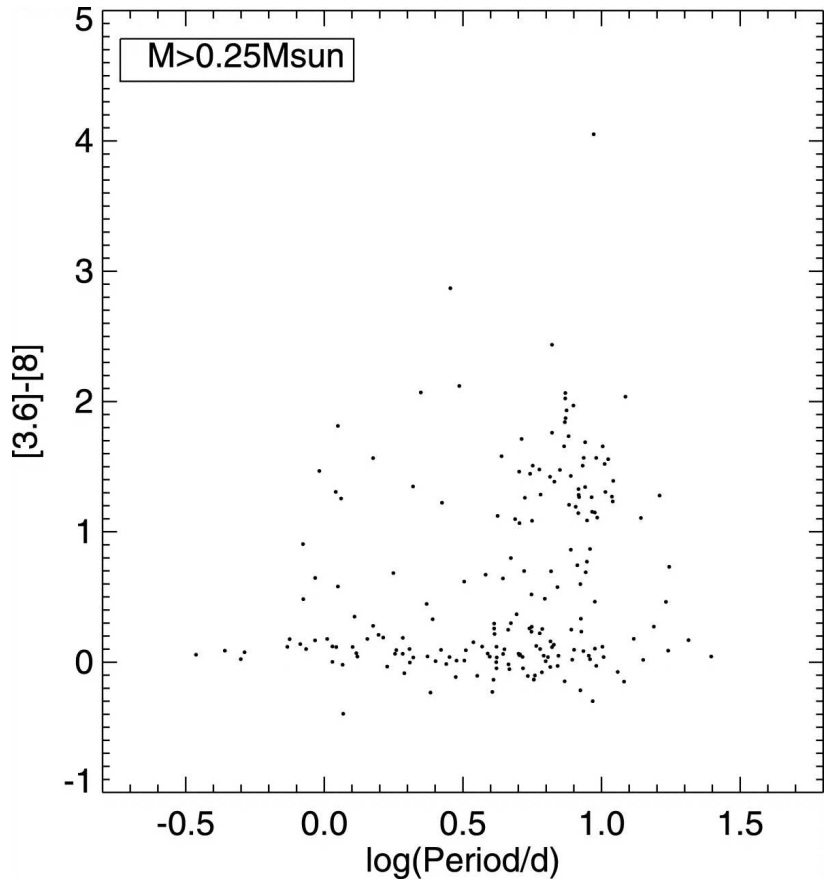}
  \caption{Spitzer [3.6]-[8.0] color excess as a function of
  rotational period for Orion low mass stars
  (M$\geq$0.25~M$_\odot$). Stars with disks have
  [3.6]-[8.0]$\geq$1. From Rebull \etal\ (2006)}\label{rebull}
\end{figure}

Keeping in mind the limitations associated to the various accretion
diagnostics, a number of recent studies have investigated the
accretion-rotation connection for relatively large samples of young
stars with known rotational periods. Lamm \etal\ (2005) used
H$_\alpha$ equivalent width to distinguish between accreting and non
accreting T Tauri stars in NGC~2264 and found the former to be on
average slower rotators than the latter. Similarly, Rebull et
al. (2006) found that Orion low mass stars with longer rotational
periods were more likely than those with short periods to exhibit a
continuum mid-IR excess indicative of disks. At odds with these
results, Cieza \& Baliber (2006) failed to find any correlation
between accretion and rotation for IC~348 low mass stars, even though
they used the same accretion diagnostics as Rebull \etal\ (2006).

These conflicting results on whether or not a relationship exists
between rotation and accretion in young stars may have a number of
causes. Statistical robustness is still an issue. Rebull \etal\ (2004)
demonstrated from Monte Carlo simulations that samples of at least 400
stars per mass bin and a perfect knowledge of their accretion status
would be needed to distinguish between the period distribution of
accreting and non accreting stars. Current samples are still about 10
times smaller per mass bin and ambiguity remains for a fraction of
stars regarding their accretion status.

Rebull \etal's (2006) results for Orion low mass stars are shown in
Figure~\ref{rebull}. Four groups of stars can be distinguished in the
mid-IR excess versus rotational period diagram, of which only 2 fulfill the
expectations of accretion-regulated angular momentum evolution: slow
rotators with strong IR excess, i.e., stars still actively accreting from
their disk and thus prevented from spinning up, and fast rotators with no
mid-IR excess, i.e., diskless stars free to spin up as they contract. The
third group consists of only a few stars with short periods which exhibit a
significant mid-IR excess. Owing to the small number of stars in this
group, it is conceivable that they represent a transient state of fast
rotation, on a timescale of 0.1~Myr, in spite of disk
accretion. Interestingly, this group might hint at a discontinuous PMS
braking process (e.g Popham 1996). The fourth group consists of a
significant fraction of slow rotators with no mid-IR excess. Thus, among
stars without IR excess, about half have periods longer than 5 days.

This latter group is the most puzzling as these stars lack evidence
for a disk and yet rotate slowly. They are often interpreted as having
dissipated their disk recently, and thus have not had time yet to spin
up. However, similar groups of slowly rotating and apparently non
accreting stars are observed in other star forming regions, over the
age range from $\leq$1~Myr (e.g. ONC) to $\sim$5~Myr
(e.g. Taurus). The spin up rate scales as R$_{star}^2$ during the
first few Myr, as long as the star remains fully convective, and
increases later on as the radiative core develops. Thus, assuming an
average initial period of 8~days at 1~Myr, a star would spin up to
periods shorter than 5~days in about 1~Myr. One thus has to assume
that, in each of the observed regions, a significant fraction of stars
($\sim$30\%) were released from their disk less than a Myr ago over
the age range 1-5~Myr.

Overall, signatures of the accretion-rotation connection, as expected
from accretion-regulated angular momentum evolution, have been
recently reported. However, some intriguing results remain. In
particular, a significant subgroup of apparently non accreting stars
have long periods, which does not fit the accretion-regulated angular
momentum scenario. Clearly, further characterization of these slow
rotators is needed to assess their actual accretion status. Also,
additional Monte Carlo simulations would help to clarify the
interpretation of the accretion-rotation diagrams over PMS evolution
timescales.

\section{Models of angular momentum evolution and disk lifetimes}

\begin{figure}
\centering \includegraphics[width=1.0\textwidth]{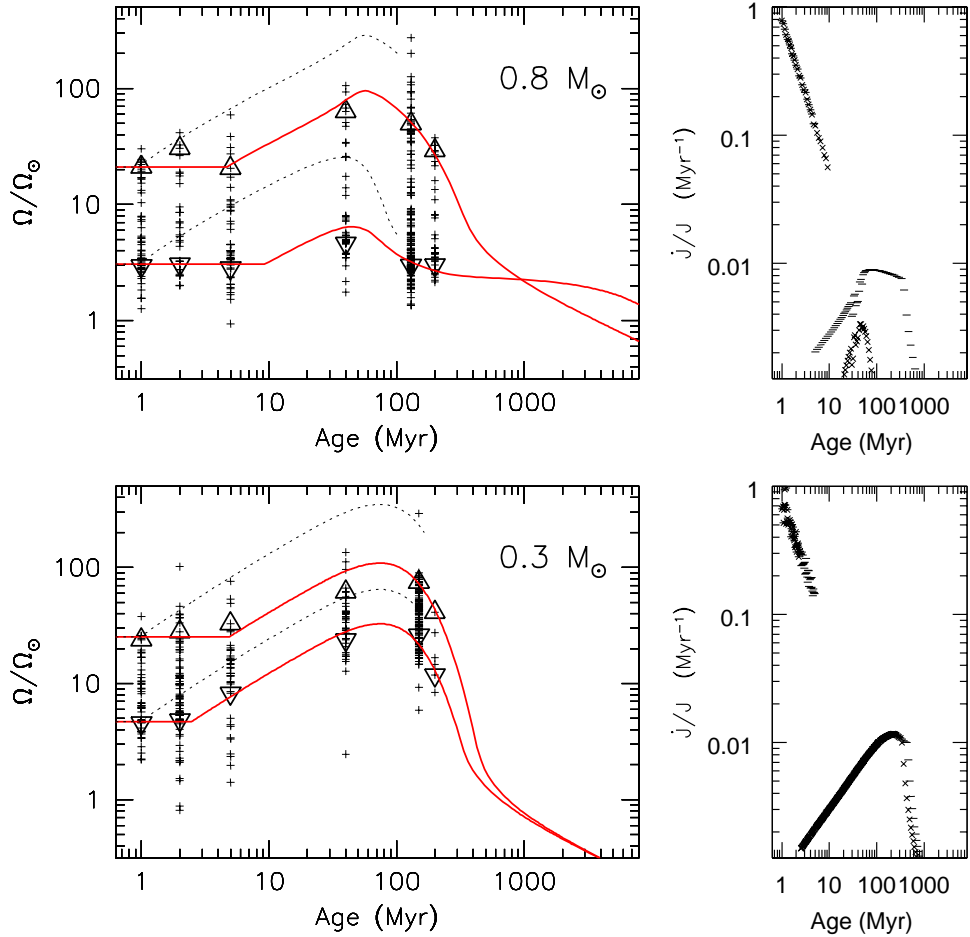}
  \caption{Angular momentum evolution models for 0.8~M$_\odot$ and
  0.3~M$_\odot$ stars. {\it Left~:} The distribution of rotational
  periods for stars in the mass range 0.7-0.9~M$_\odot$ ({\it top})
  and 0.20-0.35~M$_\odot$ ({\it bottom}) are shown by crosses. Large
  triangles indicate the 25th and 90th percentiles of the distribution
  in each cluster. Models for slow and fast rotators are shown by
  solid lines.  The initial rotational periods at 1~Myr are 1.2 and
  8.3~days for 0.8~M$_\odot$ stars ({\it top}) and 1.0 and 5.4~days
  for 0.3~M$_\odot$ stars ({\it bottom}). Disk locking timescales
  range from 2.5 to 10~Myr. Once the star is eventually released from
  its disk, solar-type winds are the only source of angular momentum
  loss. Dotted lines show similar models over the first 200~Myr for
  stars released from their disk at 1~Myr. See Bouvier, Forestini \&
  Allain (1997) and Allain (1998) for the details of the models. {\it
  Right~:} Angular momentum loss rate (per Myr) for the models shown
  in the left panels. In order to evolve at constant angular velocity
  during the first 5-10~Myr in spite of contraction, the star must
  lose, on average, about a third of its angular momentum per Myr. The
  angular momentum loss rate from ``disk locking'' is over a hundred
  times more efficient than solar-type winds to brake the star during
  the early PMS. }
\label{models}
\end{figure}

Models of angular momentum evolution during the pre-main sequence (PMS) and
on the zero-age main sequence (ZAMS) have been developped in the mid- and
late-90's and have not much progressed since then. These models rely on 2
processes to remove angular momentum from the star : i) the so-called
``disk-locking'', which assumes extremely efficient angular momentum
removal from the star as long as it magnetically interacts with its disk,
thus forcing the star to evolve at constant angular velocity in spite of
rapid contraction (Collier Cameron, Campbell \& Quaintrell 1995; Bouvier,
Allain \& Forestini 1997), ii) a magnetized solar-type stellar wind which
extracts angular momentum from the central star, acting simultaneously as
disk locking although at a much weaker rate, and whose efficiency saturates
at high velocity (Kawaler 1988; Keppens, McGregor \& Charbonneau
1995). Clearly, these models still lack a real physical description of the
angular momentum removal processes and use instead semi-empirical
parametrized braking laws. In spite of this limitation, this class of
models have been reasonably successful in reproducing the global trends
seen in the rotational evolution of young low mass stars (see, e.g.,
Bouvier, Allain \& Forestini 1997; Allain 1998; Sills \etal\ 2000).

Fig.~\ref{models} shows angular momentum evolution models computed for
0.3 and 0.8~M$_\odot$ stars and compared to the most recent rotational
datasets discussed in the previous sections. The models were computed
for 2 initial periods in each mass bin, corresponding to slow and fast
rotators, respectively, as observed at 1~Myr in the ONC. For both 0.3
and 0.8~M$_\odot$, a fraction of stars appear to evolve at constant
angular velocity for a few Myr, and up to 10~Myr, before being
released from their disk. For comparison, models in which the stars
are released from their disk at 1~Myr are shown and predict velocities
much higher than observed on the ZAMS. Hence, disk locking appears to
be active for at least a few Myr in order to prevent stars from
spinning up as they contract. Once the stars are eventually released
from their disk, they spin up as they approach the ZAMS until angular
momentum losses from solar-type stellar winds become efficient enough
to brake them over a timescale of a few 100~Myr. Note that while these
models reproduce reasonably well the rotational evolution of low mass
stars during the PMS, they predict too strong angular momentum loss
from magnetized winds for very low mass stars on the MS (cf. Sills et
al. 2000).

The right panels of Fig.~\ref{models} illustrate the angular momentum loss
rate experienced by slow and fast rotators in the 2 mass bins. The most
striking feature seen in these plots is that in order to prevent the star
from spinning up as it contracts during the first few Myr, {\it about a
third of the stellar angular momentum must be removed per Myr}. Thus, a
slowly rotating 0.8~M$_\odot$ star reduces its angular momentum by a factor
of 5 between 1 and 10~Myr. Clearly, an extremely efficient brake must apply
to the star as long as it accretes from its disk, far more efficient indeed
than angular momentum losses due to solar-type winds. MHD star-disk
interaction models have to face this challenge : not only the star-disk
interaction does not spin the star up, but it actively brakes it. In other
words, it is not enough to balance positive accretion torques by negative
magnetic torques to reach a zero net flux of angular momentum onto the
star. Instead, it is mandatory that the star-disk interaction process
actually removes angular momentum from the star at a high rate, i.e. that
the net torque on the star be strongly negative, if the star is to evolve
at constant angular velocity in spite of contraction.

Finally, the comparison of models with observations suggests that the
rotational velocity of low mass PMS stars is regulated over a
timescale of a few Myr, typically from 2 to 10~Myr. Since the strong
brake is thought to result from active star-disk interaction, this
requires that disk accretion lasts over at least this
timescale. Current estimates of disk lifetimes have been obtained from
a variety of diagnostics, including near- and mid-IR excess, and
H$_\alpha$ emission line width. While the former probe both active and
passive disks, the latter provides more direct evidence for accretion
onto the star.  The disk fraction derived from near-IR excess amounts
to about 40-60\% at 1~Myr, with no disk remaining past 5~Myr
(Hillenbrand 2005). When diagnosed from the more sensitive mid-IR
excess and/or H$_\alpha$ width measurements, longer disk survival
times are obtained, with about 40-60\% of stars still surrounded by
circumstellar disks at 2~Myr, a fraction which decreases to about
10-25\% at 10~Myr (e.g. Damjanov \etal\ 2007; Lyo \& Lawson 2005;
Jayawardhana \etal\ 2006). Interestingly, a recent study of the 8~Myr
$\eta$ Cha cluster suggests a mean disk lifetime of 9~Myr for single
low mass stars while binary systems seem to dissipate their disks on a
timescale of 5~Myr (Bouwman \etal\ 2007). No primordial disks appear
to survive past 30~Myr (Gorlova \etal\ 2007).

\section{Conclusions}

New datasets now provide hundreds of rotational periods measured for low
mass and very low mass stars in young clusters over the age range from
1~Myr to 0.2~Gyr. The distribution of rotational periods shows a clear
evolution over the pre-main sequence contraction timescale. While most
stars tend to spin up as they descend their Hayashi tracks, a fraction
retain constant angular velocity for a few million years. This provides
clear evidence for a strong brake acting on the stars on a timescale of
2-10~Myr. The new data also confirm that very low mass stars
(M$\leq$0.3~M$_\odot$) tend to suffer lower angular momentum losses than
low mass ones (0.3-1.0~M$_\odot$). The early evolution at nearly constant
angular velocity implies angular momentum loss rates much larger that those
achievable by solar-type magnetized stellar winds. New empirical evidence
has been recently reported which confirms that the magnetic star-disk
interaction is indeed responsible for the braking of low mass stars at the
start of their PMS evolution. These new results challenge the ability of
current MHD star-disk interaction models to extract as much angular
momentum from the young star as needed to prevent it from spinning up in
spite of accretion and contraction. On the observational side, a critical
time step still to be sampled in order to better constrain accretion disk
lifetimes from rotational evolution is the 5-40~Myr range as most disks
appear to dissipate on these timescales.

\begin{acknowledgments}
I would like to thank Jonathan Irwin for providing some of the Monitor data
prior to publication.
\end{acknowledgments}


\end{document}